# SOME REMARKS ON AN EXPERIMENT SUGGESTING QUANTUM LIKE BEHAVIOR OF COGNITIVE ENTITIES AND FORMULATION OF AN ABSTRACT QUANTUM MECHANICAL FORMALISM TO DESCRIBE COGNITIVE ENTITY AND ITS DYNAMICS.

Abstract:
We have executed for the first time an experiment on mental obesrvables concluding that there exists equivalence (that is to say, quantum like behavior) between quantum and cognitive entities.Such result has enabled us to formulate an abstract quantum mechanical formalism that is able to describe cognitive entities and their time dynamics.

Elio Conte [+*], Orlando Todarello [-], Antonio Federici [+],

Francesco Vitiello [+], Michele Lopane [-]

[+] Department of Pharmacology and Human Physiology;
TIRES-Center for Innovative Technologies for Signal Detection and Processing, University of Bari, Italy;

[-] Department of Neurological and Psychiatric Sciences, University of Bari, Italy

(*) School of Advanced International Studies on Theoretical and non Linear Methodologies, Bari, Italy

and

Andrei Khrennikov [°]

[°] International Center for Mathematical Modeling in Physics and Cognitive Sciences,
MSI, University of Växjö, S-35195, Sweden

and
Joseph P. Zbilut

Department of Molecular Biophysics and Physiology, Rush University Medical Center,
1653W, Congress, Chicago, IL 60612 USA

## 1. Introduction

Recently [1], we performed an experiment whose result evidences, in a preliminary way, that the dynamics of mental states follows a quantum like behavior. The conclusion of the experiment seems that there is a kind of equivalence between quantum and cognitive entities. From this result, , it follows that we may use the theory of quantum mechanics to analyze the nature of cognitive entities. The aim of the present paper is to discuss some basic features of the experiment and to give first evidences on the manner in which an abstract quantum formalism may be applied in order to have an analysis of cognition.

## 2. The Foundations of the Experiment

Quantum Mechanics is a physical theory that for first time accounts for mind and for relationship between the mind and physical world. There was indication of such attitude in the theory starting with the founding fathers of quantum mechanics. Soon after its advent, there was a long correspondence in the years 1932-1958 between W. Pauli and C. G. Jung [2]. He introduced the concept of synchronicity [3], and W. Pauli devoted much consideration to it concluding that "it would be most satisfactory if physics and psiche could be seen as complementary aspects of the same reality". The Synchronicity concept states that we have a meaningful, although acausal coincidence of mental states with an objective, external event happening at the level of physical world in time and space.

A deeper understanding of the foundations of quantum mechanics has cleared further in the past few years that this theory must be understood as an expression of the unity of Nature, and cognitive sciences move progressively on this same direction. N. Bohr was significantly influenced by the psychologist W. James in the development of the principle of complementarity that is at the basic foundation of such a theory. He was well informed on some theses discussed by James in his Principles of Psychology [4], and M. Janmer in 1989 evidenced that [5] it was Jame's use of complementarity in psychology that possibly had a great influence on Bohr's subsequent formulation of this principle in quantum mechanics.

Quantum mechanics describes the link between human cognition and physical world. Generally, in analysis of quantum mechanics a crucial role is assigned to an unavoidable physical interaction between the measuring apparatus and the physical entity to be observed. According to the founding fathers of the theory such unavoidable interaction is considered to be responsible for the uncertainty principle. Actually, only when this physical interaction is considered to be profoundly linked to a "knowledge factor" and thus to the thinking being, it has a crucial role during quantum measurement. Here we outline that just in the past years a detailed discussion of some gedanken experiments, as in particular it was made by D. M. Snyder [6], lead this author to admit that the physical interaction by only is not necessary to effect the change in the wave function that happens during the measurement in quantum mechanics. He evidenced that the "knowledge factor" is linked to the change in the wave function and not the physical interaction between the physical existent measured and the measuring instrument.

The thesis is thus that quantum mechanics is fundamentally a theory concerned with the role of the "knowledge factor" in the physical world. Of course, the "knowledge factor" is the primary manifestation of cognition that is expression of existing cognitive entities of mind. Consequently, cognition, thinking entity, and physical world are linked in quantum mechanics. As previously said, recently [1] we gave a preliminary experimental confirmation on such possible quantum like behavior of mental states in the cognition process.

There are profound theoretical elaborations that also support evidence of quantum foundations in the same principle of information coding by brain. It seems to be based on the golden mean. In the past psychologists often claimed memory span to be the missing link between psychometric intelligence and cognition. Pascual-Leone [7] in 1970 applied Bose-Einstein statistics to learning experiments obtaining a fit between predicted and tested span. H.Weiss and V. Weiss [8], multiplying span by mental speed, using entropy treatment for bosons, obtained the same result. Such authors obtained it from EEG considering span as the quantum number n of a harmonic oscillator. They arrived to the conclusion to consider the metric of brain waves as a superposition of n harmonics times $2\phi$ where half of the fundamental is the golden mean $\phi$ (1.618) as the point of resonance. Such wave packets scaled in powers of the golden mean have to be understood as numbers with directions, where bifurcations occur at the edge of chaos. In particular, such authors discussed similarities with El Naschie's theory [9] for high energy particle's physics.

## 3. Determinism and Indeterminism at Cognitive Level.

In previous papers [10], M. Zak showed that quantum mechanics may be derived from Newton's classical physics if an unnecessary mathematical restriction, the Lipschitz condition, is removed from the mathematical formalism. Non separability and non locality quantum like features were also discussed. M Zak [10] also studied the particular role of Non-Lipschitz dynamics for neural net modelling and J.P Zbilut [11] studied in particular the complexity of physiological systems whose dynamics, relaxing Lipschitz conditions, cannot be modeled by the usual deterministic or chaotic deterministic equations of motion. Rather recently some of us [12] produced a further effort analyzing physiological systems at the experimental and theoretical levels giving further evidence and support to Zbilut and Zak's non Lipschitz approach. In substance, quantum and quantum like models indicate the general basic role of indeterminism in Nature. Instead in classical physics the determinism was indicated as a basic universal law. Such philosophy was guided from the idea of Laplace's prime intelligence that marked deterministic vision as ascribed in an universal manner to Nature. Fortunately, quantum mechanics started to advise us on the importance of quantum mechanical probabilities that are ontologically present in reality itself. In many of the arguments that we will develop in the present paper we will account mainly for such basic questions that were posed from the founding fathers of the theory and also recently deepened [see 13 and 15]. Let us analyze in detail the problem of probabilities.

Also in the classical approach to physics can be explained still by a "knowledge factor". They are due to our lack of knowledge about an assumed rigid and universal determinism underlying our reality. Quantum probabilities are different. The core of the difference is that in quantum probability we have to account for a particular process that we may call potentiality- actualization. To explain, consider at an operative level an act during a measurement process; it takes place in order to ascertain and to quantify the basic features of some considered property for a given entity. If we assume rigid determinism, the measurement process has only an adaptive function. In this case, in fact, all the properties of a considered entity never exhibits potentiality, they are assumed to have always definite values before of the measurements and such values represent the automatic outcome of such measurements. We have not potentiality and we have not an effective actualization during a measurement. In the case of quantum probabilities we have instead to refer ourselves to the model potentiality- actualization previously indicated. Here, probabilities are related to the process of actualization of one among different ontological possibilities, ontological potentialities that determine the result of the measurement. In this case, the basic factor that we must account, is represented by the term context. Let us see how in detail the differences between classical and quantum-ontological probabilities arise.

Consider a dichotomous variable A =+,- . A participant is asked to answer A=+(yes) or A=- (not) to the request if he has read or not the author Dante Alighieri. This is a case of a measurement process in which the aim is to ascertain the value of A (A= yes or not). This is a classical framework with classical probabilities. Here the property has a definite value also before the measurement since the participant has a predefined opinion and answer. The measurement has here a predefined outcome. Potentiality does not exist, the context has not influence, the measurement is an automatic recording of the outcome.

Let us consider instead the case in which the following question is posed to a man: do you like this woman?, and the photograph of a previously unknown woman appears. In this case the participant has not an opinion previously established, the property here has not a definite value before the measurement, one answer of the participant will be actualized among two possible and potential answers that will formed at cognitive level only at the moment in which the question will be posed. We are no longer in a classical statistical framework. The reason is that in this case the answers A=+ or A=- will be strongly dependent on the state of the participant at the moment the question is posed but it will be also extraordinary dependent on the context in which the question is posed. In particular, the answers A=+ or A=- do not preexist but will be formed (actualized) only at the moment of the interaction between the participant and the question starting soon the participant with the two quantum like superposition of potentialities A=+ and A=- established at his cognitive level. The term context is thus determinant. Quantum probabilities, owing to the presence of the intrinsic indetermination that is linked to the potentialities and to the passage from potentiality to actualization, in brief, owing to the presence of those ontological probabilities previously discussed , result to be strong dependent from the context. Such features were discussed in [13] and M. Zak and J. P. Zbilut [10, 11] recalled the particular role of the context in systems relaxing Lipschitz conditions and thus exhibiting intrinsic indetermination. In such papers the role of the context was ascribed to the fluctuations as induced by noise. These authors explained, in fact, as an alternative approach to the traditional deterministic vision, that many oscillating phenomena in biological systems are found to be not strictly periodic but rather to be punctuated by apparent pauses having their mathematical counterparts in repelling points violating Lipschitz conditions. It was shown that in such case the role of the context is determinant in selecting one of the potentially possible, future trajectories of the system. .In some papers M. Zak and J. P. Zbilut [14] explained that such systems, containing noise and violating Lipschitz conditions, must be considered like acting and thinking systems since action is represented from transition from one equilibrium point to another and thinking is the decision making process that is based on the sign (i.e., positive or negative) of noise only at the instant in which the system reaches non Lipschitz singularities. Thus the role of context is essential in analysis of quantum and quantum like systems.

Until here, we have employed three basic terms: potentiality-actualization and context. D. Aerts and collaborators discussed in detail such basic features and their link with quantum mechanics[13]: in the formulation, such terms configure a new framework also in analysis of cognitive entities. Potentiality seems to correspond to a pure cognitive feature, actualization seems to correspond more to an act of emergent creation and, finally, context seems to be the basic interaction that is realized in the considered dynamics. It follows that, in order to give rise to stimulus-respons reflexes, our conscious experience is filtered on one hand through innate categories and, on the other hand, through aquired conceptual categories that are not fixed but they are actualized in dependence from the context . In this manner the perceived stimuli, the aquired memories and the concepts compete to structure the entities of our cognitive sphere. The context dependence of the conceptual categories assures that cognitve entity has properties that may be still evaluated and measured through cognitive processes as perception or other forms of human interactions.This was precisely the object of the experiment we have performed.

## 4. Illustration of the Performed Experiment

It has been argued in the previous sections that quantum probabilities are fundamentally different from classical probability approach since the first one involves the basic terms of potentaility of state, and of actualization that must be intended essentially in terms of emergency and creation of a state and recalls the basic role of the context. In detail, the core of the differnce resides in the fact that in quantum probability we are dealing with an act of emergency, of creation, that is the actualization of a certain property that follows to a previous potentiality and is resolved during the measurement. This is an indeterministic passage in which the context plays a central role. All such terms as potentiality, actualization and context have a great role in characterizing the dynamics of quantum entities as well as we have attempted to evidence the same decisive role in the case of cognitive entities. In order to perform experiments, we should be able to realize probability mathematical elaborations that should be idoneous to discriminate between the classical probability cases in which the context has not influences and quantum ones in which instead the context has a decisive role. With such formulas applied to cognitive entities we could analyze and conclude for classical or quantum dynamics of cognitive entities. This is what, in effect, was performed in our experiment. One of us has dedicated very much of his work to the problem of the context in probabilities [15], rather recently, he was able to treat formally the role of context in probabilities in a definitive manner. He was able to obtain two different probabilistic expressions to discriminate between situations having an active role of the context and situations having instead an ininfluent role for the context. Since these two formulas, corresponding respectively to the case of quantum like and classical behaviors, are immediately testable by experiments, it results very interesting to utilize them in some experiments involving a cognition process in manner to definitively test if, as repeatedly formulated in the present paper, quantum like behavior may be ascribed to preside over mind entities, mental states and cognition entities.

Following the elaboration given in [15], we will be interested to the so called trigonometric transformations corresponding to context-transitions that induce statistical deviations of relatively small magnitudes and that are negligible in the cases of phenomena supported from classical physics. Here the most interesting case is that one of quantum-like violation of classical formula of total probability. Quantum violation of the classical formula of total probability, based on classical formula for conditional probabilities:

$$p(A=x) = p(B=+)p(A=x/B=+) + p(B=-)p(A=x/B=-), \qquad (1)$$

where $x=+,-$, with A and B dichotomous random variables assuming $A=+,-$ and $B=+,-$, gives

$$p(A=x) = p(B=+)p(A=x/B=+) + p(B=-)p(A=x/B=-) +$$

$$+ 2\sqrt{p(B=+)p(A=x/B=+)p(B=-)p(A=x/B=-)} \cos\vartheta(x) \qquad (2)$$

where $\vartheta(+)$ and $\vartheta(-)$ are angles of phases, and give measure of the happened trigonometric transformation. We say that systems following the (2) are context dependent and exhibit quantum-like behavior. Instead, this does not happens for system following the (1). We must outline that the formalism given in [15], was developed without the application of restrictions regarding its range of validity and thus it has an absolute, general validity. In conclusion, we may apply such quantum-like formalism not only to the description of non deterministic micro phenomena but also to various phenomena outside the micro wold. One such possibility is to apply quantum-like formalism to describe statistical experiments with cognitive systems.

Let $A = +,-$ and $B = +,-$ be two dichotomous mental observables. They are assumed to be two different kinds of cognitive tasks. We prepare an ensemble E of cognitive systems, that is to say, we select an ensemble of human beings . Then, on E we perform measurements of $A = +,-$ and we obtain ensemble probabilities

$p(A=+)$ and $p(A=-) = 1 – p(A=+)$

so that $p(A=+)$ is the probability to get the result (+) under the measurement on cognitive system belonging to E, and similarly $p(A=-)$ is the probability to get the result (-).

The same result may be obtained if we perform measurement by the dichotomous mental observable $B= +, -$ obtaining $P(B=+)$ and $p(B=-) = 1 – p(B=+)$.

The next step is to prepare two ensembles $E_1$ and $E_2$ of cognitive systems having the states corresponding to $B=+$ and $B=-$. We may perform now the measurement of the A mental observable and thus obtaining

$p(A=+/B=+)$ ; $p(A=-/B=+)$ ; $p(A=+/B=-)$ ; $p(A=-/B=-)$.

$P(A=i/B=j)$ represents the probability to obtain $A = i$ ($i = +,-$) having previously obtained $B = j$ ($j=+,-$). The previously mentioned classical probability theory tell us that all these probabilities result to be connected by the well known formula of total probability (1) while instead, in the case of quantum like behavior for such explored mental states, we will have that the quantum-like formula of total probability, the (2), must hold. In this second case, in opposition to the previous classical one, we will conclude that the cognitive process employed in the dynamics of formation of the mental states in the case of the ensemble E, submitted to measurements of mental observables $A=+, -$ and $B =+, -$ is a context dependent (mental) dynamics that consequently accords to quantum like behavior.

Let us explain the experiment in detail. It is well known that, starting with 1912 [16], Gestalt moved a devastating attack against the structuralism formulations of perception in psychology. Classical structuralism theory of perception was based on a reductionistic and mechanicistic conception that was assumed to regulate the mechanism of perception. There exists for any perception a set of elementary defining features that are at the same time singly necessary and jointly sufficient in order to characterize perception also in cases of more complex conditions. Gestalt approach introduced instead an holistic new approach showing that the whole perception behavior of complex images never results to be reduced to the simple identification and sum of elementary defining features defined in the framework of our experience.

During 1920's and 1930's Gestalt psychology dominated the study of perception. Its aim was to identify the natural units of perception explaining it in a revised picture on the manner in which the nervous system works. Gestalt main contributions have provided to day some understanding elements of perception through the systematic investigation of some fascinating features as the causes of optical illusions, the manner in which the space around an object is involved in the perception of the object itself, and, finally, the manner in which ambiguity plays a role in the identification of the basic laws of the perception. In particular, Gestalt psychology also gave important contributions on the question to establish how it is that sometimes we see movements when the object we are looking at, is not really moving. As we know, when we look at something we never see just the thing we look at. We see it in relation to its surroundings. An object is seen against its background . In each case we distinguish between the figure , the object or the shape and the space surrounding it that we call background or ground. The psychologist E. Rubin [17] was the first to systematically investigate this phenomenon, and he found that it was possible to identify any well-marked area of the visual field as the figure and leaving the rest as the ground. However, there are cases in which the figure and the ground may fluctuate and one is faced to consider the dark part as the figure and the light part as the ground, and viceversa. Only a probabilistic answer may be given on a selected set of subjects that will tend to respond on the basis of subjective and context dependent factors. The importance of figure-ground relationship lies in the fact that this early work of Rubin represented the starting point from which the Gestalt psychologists began to explain what to day are known as the organizing principles of perception. A number of organizing or grouping principles emerged from such studies of ambiguous stimuli. Three identified principles may be expressed as similarity, closure and proximity. Gestalt psychologists attempted to extend their work also at a more physiological level postulating an existing strong connection between the sphere of the experience and the physiology of the system by admitting the well known principle of isomorphism. This principle establishes that the subjective experience of human being as well as the corresponding nervous event have substantially the same similar structure. Rather recently [18], A. Keil, M. Müller, W.J. Ray, T. Gruber, T. Elbert presented evidence of a neural correlation of the differential treatment of figure and ground by the brain. The effect of figure/ground assignment was observed even in the earliest portions of cell response, suggesting an intimate coupling between shape selectivity and figure/ground segregation. These new physiological findings resulted in a satisfactory accord with the perceptual effects that were described by E. Rubin.

In the experiment, we examined subjects by tests A and B in order to test quantum like behavior as predicted by (2). For test A and B we used the ambiguity figures of Fig. 1 as they were largely employed in Gestalt studies. The reasons to use such ambiguity tests here to analyze quantum like behavior in perception, may be summarized as it follows. First of all, the Gestalt approach was based on the fundamental acknowledgement of the importance of the context in the mechanism of perception. Quantum like behavior formulates the same basic importance and role of the context in the evolution of the considered mechanism. Finally, we have seen that in ambiguity tests, the figure and the ground may fluctuate during the mechanism of perception. Thus, consequently, a non deterministic (and this is to say … a quantum like) behavior should be involved.

Ninety eight students in medicine with about equal distribution of females and males, aged between 19 and 22 years, from our University , provided informed consent to participate in the experiment.. In the first experiment a group of fifty three students was subjected in part to test A ( presentation of only test A) and in part to tests B and A (presentation of test B and soon after presentation of test A with prefixed time separation of about two seconds between the two tests). The same procedure was employed in the second and the third experiments for groups of twenty four and twenty one students, respectively. All the students of each group were submitted simultaneously to test A or to test B followed by test A. Ambiguity figures of tests A or B followed by A appeared on a large screen by a time of only three seconds and simultaneously it was asked to students to bar on a previously prepared personal schedule their decision to retain figures to be equal or not. Submission to students of test B soon after followed by test A had the finality to evaluate as the perception of the first image (test B) could alter perception of the subsequent image (test A). All the experiments were computer assisted and in each phase of the experimentation the following probabilities were calculated

$p(A=+), p(A=-), p(B=+), p(B=-), p(A=+/B=+), p(A=-/B=+), p(A=+/B=-), p(A=-/B=-)$

and than a statistical analysis of the results was performed in order to ascertain if the (1) or instead the (2) occurred in the case of our measurements of mental observables as they were performed by us with Tests A, B, and A/B.

The first experimentation gave the following results

Test A : $p(A=+) = 0.6923$ ; $p(A=-) = 0.3077$,                                   (3)

Test B : $p(B=+) = 0.9259$ ; $p(B=-) = 0.0741$ ;

Test A/B : $p(A=+/B=+) = 0.68$ ; $p(A=-/B=+) = 0.32$ ,

$p(A=+/B=-) = 0.5$; $p(A=-/B=-) = 0.5$

The calculation of conditional probability gave the following result with regard to $p(A=+)$

$p(B=+) p(A=+/B=+) + p(B=-) p(A=+/B=-) = 0.6666$               (4)

The second experimentation gave the following results

Test A : p(A=+) = 0.5714 ;  p(A=-) = 0.4286 ,                    (5)
Test  B: p(B=+) = 1.0000 ;   p(B=-) =0.0000 ;
Test A/B : p(A=+/B=+) = 0.7000 ; p(A=-/B=+)0.3000 ;
p(A=+/B/-) = 0.000 ;   p(A=-/B=-) = 0.0000 .
The calculation of the conditional probability gave the following result with regard to p(A=+)
p(B=+)p(A=+/B=+) + p(B=-) p(A=+/B=-) = 0.7             (6)
Finally, the third experimentation gave the following results
Test A : p(A=+) =0.4545 ;   p(A=-) = 0.5455 ,                    (7)
Test B:  p(B=+) =0.7000 ;   p(B=-) = 0.3000 ;
Test A/B : p(A=+/B=+) = 0.4286 ;  p(A=-/B=+) = 0.5714; p(A=+/B=-) = 1 , p(A=-/B=-) = 0
The calculation of the conditional probability with regard to p(A=+) gave the following result
P(B=+) p(A=+/B=+) + p(B=-) p(A=+/B=-) = 0.6000                (8)

It is seen that the mean value of p(A=+) resulted in p(A=+) = 0.5727 ± 0.1189 with regard to the Test A and calculated by using the (3), the (5), and the (7) while instead a mean value of 0.6556 ± 0.0509 resulted for p(A=+) when calculated with regard to the Test A/B and thus using the (4), the (6), and the (8). The two obtained mean values were different and thus give evidence for the presence of quantum like behavior in measurements of cognitive mental states as they were performed by testing mental observables by Tests A, B, and A/B. The use of Student's t-Test demonstrated that we had no more than a 0.30 probability that the obtained differences between the two estimated values of p(A=+) by Test A and by Test A/B were produced by chance.

As final step we may proceed now to calculate $\cos\vartheta(x)$ as given in the (2) and representing, as we know, a trigonometric measure of transformation of probabilities that are generated by transition from one context (the case of Test A) to another context (the case of Test A/B) . In the case of our experimentation we obtained $\cos\vartheta(+) = -0.2285$, $\vartheta(+) = 1.8013$ and $\cos\vartheta(-) = 0.0438$, $\vartheta(-) = 1.527$ that are quite satisfactory phase results in order to admit quantum like behavior for the investigated mental states.

On the basis of the illustrated results ,we concluded that we had a preliminary evidence of existing quantum like behavior in the dynamics of mental states. Luckily, we were able to capture mental conditions of subjects in which the context influenced decision in an essential way. We had equivalence between quantum and cognitive entities.

## 5. Application of the Abstract Quantum Formalism to Analysis of Cognitive Entities

The performed experiment establishes that we have a true quantum like behavior of cognitive entities. A direct consequence of the results of the experiment is that cognitive entities as well as quantum entities experience an high contextual nature. As well as quantum entities are influenced by the usual physical act of measurement, also cognitive entities are influenced by the act of measurement. In the case of cognitive entities, the measurement is characterized by the cognitive interaction. According to the fact that we have a nonfixed character of knowledge, it follows that also cognitive states must be intended not having states with fixed character. This is well as the case of quantum entity during a measurement, whose possibility of identifying its ontology will depend on the universality of the experimentation that will be employed during its measurement. It follows that the theory of quantum mechanics can be used fully in analyzing the nature of cognitive entities.  Let us consider that , in order to explore the context dependence of cognitive entities, we have to introduce the explicit dependence of the truth and falsehood of a given sentence on the cognitive interaction with the cognitive person. Reading a sentence, or "making a sentence true or false" we will "perform a measurement " on the sentence This means that we start with a state of potentiality: a sentence  is in general not true  and not false. We have a superposition of both such possibilities. During the act of the measurement, the state of the sentence changes in such a way that  it becomes true or false. This state of potentiality, neither true nor false, will be called a superposition of potentialities, and it will be followed by actualization , that is an act of creation in which the new emergent state (sentence true or false) will take place during the cognitive measurement. Thus, in conclusion, before the cognitive measurement starts, and this is to say: when we, human thinking beings, start to interact with the sentence, i.e. reading it, the sentence is in a potential state in the sense that it is not neither true nor false for us, and we call it a superposition state. As consequence of the cognitive measurement, finally the sentence becomes or true or false in consequence of the interaction subjected by the thinking Human subject with the sentence in a particular given context that will influence the result of the choice by a non deterministic dynamics.

Cognitive entity will be expressed by a non determinate state of truth and falsehood in the following terms

$$\psi = c_{true}\varphi_{true} + c_{false}\varphi_{false} \quad (9)$$

where $\varphi_{true}$ is the "state true"  and $\varphi_{false}$ is the "state false" of the cognitive entity. $c_{true}$ and $c_{false}$ are pondering factors whose square modulus will give the statistical probabilities of finding the cognitive entity in the state "true" or "false" respectively, after the cognitive measurement with its contextual influence will happen. In brief, we will have that

potentiality = $\left( \psi = c_{true}\varphi_{true} + c_{false}\varphi_{false} \right) + \left( \text{context-cognitive-measurement} \right)$

→ or $c_{true}\varphi_{true}$ or $c_{false}\varphi_{false}$ (emergent creation)  (10)

where $|c_{true}|^2$ will represent the probability that the true state will be actualized in consequence of the context-cognitive measurement, while $|c_{false}|^2$ will represent instead the probability that the false state will be actualized in consequence of the same measurement. Here the ontological nature of such probabilities is evidenced. An unequivocal result, true/false, cannot be obtained in consequence of context-cognitive measurement since a superposition of potentialities is present with different pondering factors.. An unequivocal response true or false may be obtained only statistically when the potential superposition is actualized.This is the quantm like behavior. In the classical one, instead, only one outcome may be obtained as result of the cognitive measurement and precisely that one previously defined. In this case cognitive measurement has only an adaptive role of recording a predefined answer and the context has not influence.

In this manner it is evidenced the importance of the operation that we are attempting in the present paper. Based on the equivalence between cognitive and quantum entities that we evince on the basis of the previous experiment, we are making direct use of an abstract quantum formalism applied to cognitive entity and this means that we are becoming able to delineate the basic features of a cognitive entity using quantum language and formalism.There is more. On the basis of the (1) and (2) we are for the first time the possibility to make accurate quantum mechanical elaboration on the cognitive entity taken in consideration. As example, the numerical results obtained on the basis of the previous experiment gives immediately the opportunity to delinate basic features of cognitive entities that never were known in the past. Let us follow our appliction in detail.We may introduce the complex quantum amplitude that will represent the state of our cognitive entity expressed in relation to some selected mental obeservable. Let us admit that we selected the mental observable A pertaining to a given cognitive entity. Let us admit also that A,as previously fixed, could assume only two possible values (A=+,- ). Such complex quantum amplitude will be given through the following formula

$|\phi(\pm)|^2 = P(A = \pm)$   (11)

and it will represent the state of our cognitive entity in relation to the considered mental observable A. The relevant feature of the present quantum elaboration is linked to the fact that we may be able to perform direct experimentations on some mental observable A, B, C,….. and consequently, using the previous equation, we may be able to deduce the quantum complex amplitude, that is to say .,. the quantm state of the cognitive entity in relation to the mental obesrvations that will be executed on A or B or C Thus, for the first time, we arrive to the condition to attribute direct to mental functions an abstracted quantum formalization.Just our performed experiment may give us the opportunity to express in detail a methodological indication on the manner in which each researcher will be free to apply such new elaborationin the course of future experiments. We may briefly reconsider the case of the experiment that we performed experiment, showing how to arrive to write quantum complex amplitudes and thus to give a quantum characterization of the state of the cognitive entity that was employed in such experiment. Let us consider in detail that we started by the (2) that was confirmed to exist in the case of our experimentation. As we indicated previously, we arrived to calculate two different values for $\cos\vartheta(+)$ and $\cos\vartheta(-)$ whose meaning is now well clear.In the case of our experimentation we obtained $\cos\vartheta(+) = -0.2285$, $\vartheta(+) = 1.8013$ and $\cos\vartheta(-) = 0.0438$, $\vartheta(-) = 1.527$ that are quite satisfactory phase results in order to admit quantum like behavior for the investigated cognitive entity. As final step, we may now proceed by a detailed calculation of the quantum like model of the mental states of the cognitive entity as they were characterized during the experimentation.

By using the obtained data, we can write a quantum-like wave function $\phi = \phi_S$ of the mental state S (of the group of students participated in the experiment). The general formula for representation by the quantum complex amplitude of the cognitive entity was given in [12], and it may be rewritten here in the following terms

$\phi(x) = \sqrt{P(B=+)\,P(A = x / B = +)} + e^{i\theta(x)} \sqrt{P(B=-)\,P(A = x / B = -)}$   (12)

The $\phi$ is a function from the range of values { +, -} of the mental observable A to the field of complex numbers. Since the A may assume only two values, such function can be represented by two dimensional vectors with complex coordinates. Our experimental data will give

$\phi(+) = \sqrt{0.8753 \times 0.6029} + e^{i\theta(+)} \sqrt{0.1247 \times 0.5} \approx 0.7193 + i\, 0.2431$   (13)

and

$\phi(-) = \sqrt{0.8753 \times 0.3971} + e^{i\theta(-)} \sqrt{0.1247 \times 0.5} \approx 0.5999 + i\, 0.2494$   (14)

Thus, in conclusion, let us see how our elaboration enables to start from experimental results obtained in the course of a programmed experiment to arrive to write explicitly the states of the involved cognitive entities by mathematical functions. They are now in a so explicit form that we are able to analyze and to test them in all the required experimental conditions.

Entering briefly in the field of very technical notations, we may illustrate here that, by performing the same experiment for every group of people having some special mental state G we can calculate the wave function $\phi_G$ giving a quantum-like representation of the G. Thus there exists the map J mapping mental states into quantum-like wave functions. Such a mapping provides a mathematical representation of mental functions. A mental state is too complex object to provide its complete mathematical description, but we can formalize mathematically some features of a mental state by using the quantum-like representation. We remark that every quantum-like representation induce the huge reduction of information. In particular, the wave function $\phi_S$ calculated in our experiment gives a very rough representation of the mental state S. The $\phi_S$ contains just information on ability of students that were employed in the experiment, to percept contexts on the A and B-pictures.

On the space of complex functions we introduce the structure of a Hilbert space H with the aid of the scalar product

$$(\phi, \psi) = \phi(+) \overline{\psi}(+) + \phi(-) \overline{\psi}(-) \tag{15}$$

Thus J maps the set of mental states into the H. The mental observable A can be represented by the multiplication operator in H:

$$\hat{A} \phi(x) = x \phi(x) \ ; \ x = \pm \tag{16}$$

We see [11] that the mean value of the mental observable A in the mental state S can be calculated by using the Hilbert space representation

$$E_S A = (\hat{A} \phi, \phi) \tag{17}$$

In our concrete experiment by using experimental data, we have

$$E_S A = P(A = +) - P(A = -) = 0.1454 \tag{18}$$

The same result gives our quantum-like model

$$(\hat{A} \phi_S, \phi_S) = | \sqrt{0.8753 \times 0.6029} + e^{i\theta(+)} \sqrt{0.1247 \times 0.5} |^2 - | \sqrt{0.8753 \times 0.3971} +$$
$$+ e^{i\theta(-)} \sqrt{0.1247 \times 0.5} |^2 = 0.1454 \tag{19}$$

Thus, in conclusion, we arrive at a complete quantitative representation of the basic features of the cognitive entity that was engaged in our experimentation obtaining, in particular, to calculate mean values of mental observables. Therefore, it seems that way of applying an abstracted quantum formalism to cognition, based such approach on an acknowledged equivalence evidenced at experimental level between qunutam and cognitive entities, opens new perspectives of relevant interest in studies of mental processes.

In order to complete our approach that aspires to introduce a fully application of quantum abstract formalism to the analysis of cognitive entities, we must now discuss the problem of the behavior of such entities under a dynamic profile. As repeatedly outlined in the previous sections, a cognitive entity is regulated by three basic features: we have a state of potentiality that in quantum mechanical terms is represented by an undifferentiated and indeterminate superposition state, we have a state of actualization that is the stage of emergence and creation, the state in which a given cognitive interaction, assures a given sentence to become actually true or false in subject experience and, finally, we have the action of the context that in a quantum like cognitive process, acts in fundamental terms driving indeterministically the actualization of the cognitive final acquisition. A cognitive measurement assures the passage of cognitive entity from the potential to contextual-actualized state. The problem to give a dynamical caharacterization of a cognitive entity is the question to explicit the dynamical evolution of a cognitive entity in time. We know that, when left unmeasured (that is to say…. When it is not subjected to a cognitive measurement) the entity remains statistically in its undifferentiated superposition state. Our aim is to start describing such dynamics in time. In abstract and formal terms we may say that we have to introduce a dynamical evolution operator U(t), time dependent, that acts on the initail state of the cognition entity. In the most simple case of the superposition given in $\psi$ in (9), if we indicate such state of cognitive entity by $\psi_0$ to indicate that it is related to the initial time 0, we will write that the state of the cognitive entity at any time t, will be given by

$$\psi(t) = U(t)\psi_0 \ \text{and} \ \psi_0 = \psi(t=0) \tag{20}$$

The entity starts its cycle and, if left unmeasured by some cognitive measurement, remains statistically in its undifferentiated superposition state of potentailies. If, during such dynamical evolution, some cognitive measurement will start, the dynamical evolution of the superposition state will be interrupted and a final state will be indeterministic selected among the ontological possibilities and yielded to be actualized on the basis of the intrinsic features of the entity and of its interaction and context. Before of all we would examine the nature of the dynamical time evolution expressed by the (20). Before of all, we have to attribute a physical meaning to the time t before the actualization will be performed owing to cognitive measurement and acting context. We will call it the time of the temporal evolution of the cognitive entity. Essentially, a Hamiltonian H must be constructed such that the evolution operator U(t), that must be unitary, gives $U(t) = e^{-iHt}$.

It is well known that, given a finite N-level quantum system described by the state $\psi$, its evolution is regulated according to the time dependent Schrodinger equation

$$i\hbar \frac{d\psi(t)}{dt} = H(t)\psi(t) \quad \text{with} \quad \psi(0) = \psi_0. \quad (21)$$

Let us introduce a model for the hamiltonian H(t). It is the hamiltonian of the cognitive entity. We express by $H_0$ the free hamiltonian of the cognitive entity, and we consider it as a constant- internal haminltonian component that reasumes all the basic mental, hystorical, social features of the considered entity. We than add to $H_0$ an external time varying hamiltonian, $H_1(t)$, representing the interaction of the cognitive entity with the control fields, intending by this term all the mind and also brain influences that will act on the cognitive entity during the evolution of the initial superposition state indicated by $\psi_0$. Thus, for the first time, we attempt here to give an unitary representation of a cognitive entity including in the time varying term $H_1(t)$, mental contributions as well as synchronus contributions deriving from mind-brain relation. In conclusion we write the total hamiltonian as

$$H(t) = H_0 + H_1(t) \quad (22)$$

so that the time evolution of the state of the cognitive entity will be given by the following Schrodinger equation

$$i\hbar \frac{d\psi(t)}{dt} = [H_0 + H_1(t)]\psi(t) \quad (23)$$

and $\psi(0) = \psi_0$. We have that

$$\psi(t) = U(t)\psi_0 \quad (24)$$

where U(t) pertains to the special group SU(N). We will write that

$$i\hbar \frac{dU(t)}{dt} = H(t)U(t) = [H_0 + H_1(t)]U(t) \quad \text{and} \quad U(0)=I \quad (25)$$

Let $A_1, A_2, \ldots, A_n$, ($n=N^2-1$), are skew-hermitean matrices forming a basis of Lie algebra su(N). Assuming semiclassical approximation for external acting fields $H_1(t)$, and following the previous papers developed in [19], one arrives to write the explicit expression of the hamiltonian H(t) of the cognitive entity. It is given in the following manner

$$-iH(t) = -i[H_0 + H_1(t)] = \sum_{j=1}^{n} a_j A_j + \sum_{j=1}^{n} b_j A_j \quad (26)$$

where $a_j$ and $b_j = b_j(t)$ are respectively the constant components of the free hamiltonian and the time-varying control parameters characterizing the interaction of the cognitive entity. If we introduce T, the time ordering parameter, still following in detail the previous work given in [19], we arrive also to express U(t) that will be given in the following manner

$$U(t) = T\exp(-i\int_0^t H(\tau)d\tau) = T\exp(-i\int_0^t (a_j + b_j(\tau))A_j d\tau) \quad (27)$$

that is the well known Magnus expansion [19]. Locally U(t) may be expressed by exponenetial terms as it follows [19]

$$U(t) = \exp(\gamma_1 A_1 + \gamma_2 A_2 + \ldots + \gamma_n A_n) \quad (28)$$

on the basis of the Wein-Norman formula [19]

$$\Xi(\gamma_1, \gamma_2, \ldots, \gamma_n) \begin{pmatrix} \dot{\gamma}_1 \\ \dot{\gamma}_2 \\ \ldots \\ \dot{\gamma}_n \end{pmatrix} = \begin{pmatrix} a_1 + b_1 \\ a_2 + b_2 \\ \ldots \\ a_n + b_n \end{pmatrix} \quad (29)$$

with $\Xi$ n x n matric, analytic in the variables $\gamma_i$. We have $\gamma_i(0) = 0$ and $\Xi(0) = I$, and thus it is invertible, and we obtain

$$\begin{pmatrix} \dot{\gamma}_1 \\ \dot{\gamma}_2 \\ \ldots \\ \dot{\gamma}_n \end{pmatrix} = \Xi^{-1} \begin{pmatrix} a_1 + b_1 \\ a_2 + b_2 \\ \ldots \\ a_n + b_n \end{pmatrix} \quad (30)$$

The present elaboration has reached now some central objectives that seem to be of considerable interest.
1) we have learned as to write explicitly the hamiltonian of a cognitive entity.

2) Still, we have learned how to write explicitly the time evolution unitary operator U(t) regulating the dynamic time evolution of a cognitive entity in absence ($b_j$=0) of external influences or when mental and brain influences are present.
3) Finally, we have evidenced that ,by direct experimentation conducted on cognitive entities, we may arrive to express not only the Hamiltonian H(t) of a cognitive entity and evolution operator U(t), but we may arrive also to estimate the fundamental parameters $a_j$ , $b_j$ (t) that reproduce the basic features of thecognitive entity. We may arrive to express differential equations for such paremeters and variables by the $\gamma_j$ introduced in the previous considered systems of differentail equations (29) or (30). In brief, we have arrived to express a formalism that enables to give for the first time a satisfactory characterization of the basic cognitive and neurological features that may pertain to a cognitive entity.

In the mean while we may see also how we may render still more explicit the previously obtained results.

To reach this objective we must consider a simple case of cognitive entity based on the superposition of only two states as we considered it in the (9). As we outlined, we have

$$\psi = [y_1, y_2]^T \quad \text{and} \quad |y_1|^2 + |y_2|^2 = 1 \tag{31}$$

As previously said, we have here an SU(2) unitary transformation, selecting the skew symmetric basis for SU(2). We will have that

$$e_1 = \begin{pmatrix} 0 & 1 \\ 1 & 0 \end{pmatrix} , \quad e_2 = \begin{pmatrix} 0 & -i \\ i & 0 \end{pmatrix} , \quad e_3 = \begin{pmatrix} 1 & 0 \\ 0 & -1 \end{pmatrix} \tag{32}$$

Now we will consider the following matrices

$$A_j = \frac{i}{2} e_j , \quad j = 1,2,3 \tag{33}$$

We are now in the condition to express H(t) and U(t) in some cases of interest. In this paper we will take in consideration only the most simple case, that one of fixed and constant control parameters $b_j$. In subsequent papers we will take in consideration more complex and also non linear behaviors. According to [19], the hamiltonian H of the cognitive entity will become fully linear time invariant and its exponential solution will take the following form

$$e^{t(\sum_{j=1}^{3}(a_j+b_j)A_j)} = \cos(\frac{k}{2}t)I + \frac{2}{k}\operatorname{sen}(\frac{k}{2}t)\left(\sum_{j=1}^{3}(a_j+b_j)A_j\right) \tag{34}$$

with $k = \sqrt{(a_1+b_1)^2 + (a_2+b_2)^2 + (a_3+b_3)^2}$. In matrix form it will result

$$U(t) = \begin{pmatrix} \cos\frac{k}{2}t + \frac{i}{k}\operatorname{sen}\frac{k}{2}t(a_3+b_3) & \frac{1}{k}\operatorname{sen}\frac{k}{2}t[a_2+b_2+i(a_1+b_1)] \\ \frac{1}{k}\operatorname{sen}\frac{k}{2}t[-a_2-b_2+i(a_1+b_1)] & \cos\frac{k}{2}t - \frac{i}{k}\operatorname{sen}\frac{k}{2}t(a_3+b_3) \end{pmatrix} \tag{35}$$

and, obviously, it will result to be unimodular as required .This is the matrix representation of time evolution operator for the considered cognitive entity.

Starting with this matrix representation of time evolution operator U(t), we may deduce promptly the dynamic time evolution of the state of cognitive entity at any time t writing

$$\psi(t) = U(t)\psi_0 \tag{36}$$

and rememebring that we have for $\psi_0$ the following expression

$$\psi_0 = \begin{pmatrix} c_{true} \\ c_{false} \end{pmatrix} \tag{37}$$

having assumed for the true and false states the following matrix expressions

$$\varphi_{true} = \begin{pmatrix} 1 \\ 0 \end{pmatrix} \quad \text{and} \quad \varphi_{false} = \begin{pmatrix} 0 \\ 1 \end{pmatrix} \tag{38}$$

Finally, one obtains the expression of the state $\psi(t)$ of the cognitive entity at any time

$$\psi(t) = \left[ c_{true} \left[ \cos\frac{k}{2}t + \frac{i}{k}\sen\frac{k}{2}t(a_3 + b_3) \right] + c_{false} \left[ \frac{1}{k}\sen\frac{k}{2}t[(a_2 + b_2) + i(a_1 + b_1)] \right] \right] \varphi_{true} +$$
$$\left[ c_{true} \left[ \frac{1}{k}\sen\frac{k}{2}t[i(a_1 + b_1) - (a_2 + b_2)] \right] + c_{false} \left[ \cos\frac{k}{2}t - \frac{i}{k}\sen\frac{k}{2}t(a_3 + b_3) \right] \right] \varphi_{false}$$

(39)

As consequence, the two probabilities $P_{true}(t)$ and $P_{false}(t)$ that are expected for future selection to, true or false, as consequence of cognition measurement and context influence, will be given at any time t by the following expressions

$$P_{true}(t) = (A^2 + B^2)\cos^2\frac{k}{2}t + \frac{1}{k^2}\sen^2\frac{k}{2}t(P^2 + Q^2) + \frac{\sen kt}{k}(AP + BQ)$$

and  (40)

$$P_{false}(t) = (C^2 + D^2)\cos^2\frac{k}{2}t + \frac{1}{k^2}\sen^2\frac{k}{2}t(S^2 + R^2) + \frac{\sen kt}{k}(RC + DS)$$

where
A= Re $c_{true}$, B=Im $c_{true}$, C=Re $c_{false}$, D=Im $c_{false}$,
P=-D($a_1+b_1$)+C($a_2+b_2$)-B($a_3+b_3$),
Q=C($a_1+b_1$)+D($a_2+b_2$)+A($a_3+b_3$),  (41)
R=-B($a_1+b_1$)-A($a_2+b_2$)+D($a_3+b_3$),
S=A($a_1+b_1$)-B($a_2+b_2$)-C($a_3+b_3$)

As it is seen, our initial purpose to introduce an abstract quantum formalism in order to describe the time dynamics of a cognitive entity has been now fully reached. By using proper experimentation we are now in the condition to analyze cognitive behavior in simple cases of control fields as well as in cases of more complex and non linear dynamical conditions. In any case the finality will be to analyze cognitive dynamics and its basic interactions by establishing with the experiments the correct behavior of the constant parameters $a_j$ and of the time dependent functions $b_j(t)$ that regulate the time dependent behavior of the acting control fields during the dynamics of the cognitive process.

In conclusion, we have reached all the result that we prospected starting with the present paper.

As final step, we aim now to show as the few new principles that we have introduced in the present paper, based on a found equivalence between cognitive and quantum entities, may be interesting also in analysis of more complex situations regarding the dynamics of cognitive entities. As example, by this abstract formalism, we may give a look to analysis of Gestalt that is an important field in psychology. We will not reconsider here the details of the theory but only we will limit ourselves to re outline that in Gestalt there is the crucial problem to analyze the manner in which the space around an object is involved in the perception of the object itself. We have here a very clear indication on the crucial role explained by the context during dynamical evolution of a cognitive entity. From Gestalt, it is well known in detail that in ambiguity figures we have cases in which figure and ground may fluctuate so that one is faced to consider the dark part of the image as figure and the light part as the ground. Otherwise one is attempted to select instead the light part as figure and the dark part as ground (see as example, Fig.2). We think that this an evident case of equivalence between quantum and cognitive entitieis. Here the perceptive dynamics is regulated by ontological probabilities as direct consequence of potentiality-actualization- context that we discussed as basic foundations for an application of an abstracted quantum formalism in analysis of cognitive entity.

In this case we have a starting state of potentiality. As previously outlined, the cognitive entity will be represented by an undifferentiated superposition state intrinsecally indeterministic. Following the simple case given in (9), we will have that the potential state of the cognitive entity will be given by

$$\psi_{cognitive\ entity} = \frac{1}{\sqrt{2}} \left[ \varphi_{darkpart-figure} \varphi_{lightpart-ground} - \varphi_{darkpart-ground} \varphi_{lightpart-figure} \right]$$

where $\varphi_{darkpart-figure}$, $\varphi_{lightpart-ground}$, $\varphi_{darkpart-ground}$, $\varphi_{lightpart-figure}$ represent respectively the following potential states, darkpart-figure, lightpart-ground, darkpart-ground, lightpart-figure, of the cognitive entity Actualization with the proper context will determine the final decision to select figure as dark part and light part as ground or lightpart as figure and darkpart as ground. We have an explicit case of quantum entanglement with its equivalence at the level of cognitive entities.

Thus, in conclusion, the application of such an abstract quantum formalism to cognitive entities will enable within a brief time to explore the basic fundations of mental states under a quantum mechanical perspective here including, first of all, an analysis of Bell's violation.

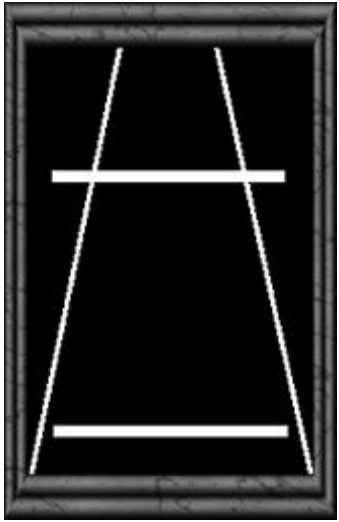
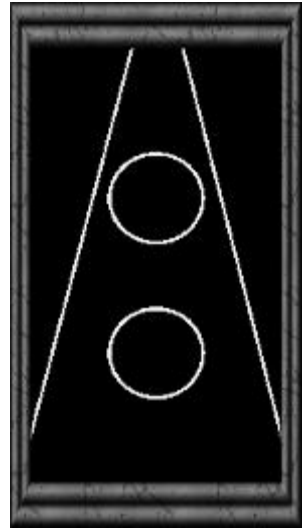

Fig.1                Test A                                Test B

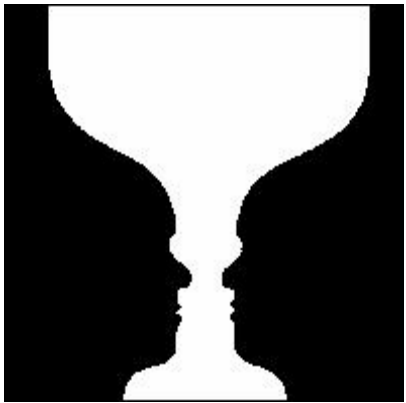

Fig. 2